\documentclass[cits,a4paper]{PoS}
\usepackage{amsmath,amssymb}
\usepackage[tight,TABTOPCAP]{subfigure}
\usepackage{graphicx}
\usepackage{grffile}
\usepackage[font=small,labelfont=bf]{caption}
\usepackage{relsize}

\newcommand{\psibar}{\overline \psi}

\newcommand{\daqhq}{\Delta \alpha_{\rm QED}^{\rm had}(Q^2)}

\graphicspath{{figs/}}

\title{Study of the hadronic contributions to the running of the QED
  coupling and the weak mixing angle}

\ShortTitle{Study of the hadronic contributions to $\daqhq$ and
  $\Delta^{\rm had}\sin^2\theta_W({Q}^2)$}

\author{

  Anthony~Francis${\,}^a$, Vera~G\"{u}lpers${\,}^b$, \speaker{Gregorio~Herdo\'iza}${\,}^c$, Hanno~Horch${\,}^b$,
  Benjamin~J\"{a}ger${\,}^{d}$, Harvey~B.~Meyer${\,}^{b,e}$, Hartmut~Wittig${\,}^{b,e}$\\
  \llap{$^a$} Department of Physics and Astronomy, York University, M3J1P3,
  Toronto, ON, Canada\\
  \llap{$^b$} PRISMA Cluster of Excellence, Institut f{\"u}r Kernphysik,\\
  Johannes Gutenberg-Universit{\"a}t, 55099 Mainz, Germany\\
  \llap{$^c$}Instituto de F\'isica Te\'orica UAM/CSIC and
  Departamento de F\'isica Te\'orica,\\
  Universidad Aut\'onoma de Madrid, Cantoblanco, E-28049
  Madrid, Spain\\
  \llap{$^d$}
  Department of Physics, College of Science, Swansea University, Swansea SA2 8PP, UK\\
  \llap{$^e$} Helmholtz Institute Mainz,
  Johannes Gutenberg-Universit{\"a}t, 55099 Mainz, Germany \\
  E-mails:
  \email{afrancis.heplat@gmail.com,\{guelpers,horch,meyerh,wittig\}@kph.uni-mainz.de,
    gregorio.herdoiza@uam.es, B.Jaeger@swansea.ac.uk} 

}

\abstract{

  The electromagnetic coupling receives significant contributions to
  its running from non-perturbative QCD effects. We present an update
  of a lattice QCD study of the Adler function and of its application
  to the determination of the running of the QED coupling. We perform
  a high-statistics computation with two flavours of O$(a)$ improved
  Wilson fermions in a large range of momentum transfer $Q^2$. The
  running of the electromagnetic coupling, including contributions
  from $u$, $d$, $s$ and $c$ valence quarks, is compared to
  phenomenological determinations at intermediate $Q^2$ values. An
  extension of this study to the determination of the hadronic
  contributions to the running of the weak mixing angle is also
  described.

}

\FullConference{The 33rd International Symposium on Lattice Field Theory\\
  14 -18 July 2015\\
  Kobe International Conference Center, Kobe, Japan*}

\begin{document}


\section{Introduction}
\label{sec:intro}

The running of the electromagnetic coupling is governed by photon
vacuum polarisation effects. When varying the virtuality, $Q^2$, of
the photon from the Thomson limit, where the QED coupling corresponds
to the fine structure constant $\alpha \equiv \alpha(Q^2=0)$, up to
the $Z$-pole mass the coupling increases by approximately
6\%~\cite{Agashe:2014kda}. Leptons and quarks contribute by roughly
the same amount to this running. The contributions from leptons,
$W$-pairs and the top quark can be accurately computed in perturbation
theory. The logarithmic corrections involving the ratio of lepton to
$Z$-boson masses are behind the relatively large leptonic
contribution. On the other hand, the running of the QED coupling also
involves energy scales where non-perturbative QCD contributions from
the lightest five quark flavours are significant.

The largest fraction of the overall uncertainty on the running of the
QED coupling is due to these low-energy hadronic effects.  While the
fine structure constant is known with a 0.3\,ppb precision, the
coupling to the $Z$-pole, $\alpha(M_Z^2)$, has a uncertainty which is
five orders of magnitude larger~\cite{Agashe:2014kda}. The prospects
for a future International Linear Collider suggest that the
uncertainty on the running of the QED coupling could become a limiting
factor in the global fit of the electroweak sector of the Standard
Model~\cite{Jegerlehner:2011mw,Baak:2014ora}.

A phenomenological approach~\cite{Davier:2010nc,Hagiwara:2011afg}
based on a dispersion relation and on the experimental measurement of
the cross-section for hadron production in $e^{+}e^{-}$-annihilation
provides a method to determine the hadronic contribution to the
running of $\alpha(Q^2)$. A similar approach is used to extract the
lowest-order hadronic contribution to the muon $g-2$. The main
difference being that in the case of the running of $\alpha(Q^2)$, the
dispersive integral is dominated by a higher-energy region.

Lattice QCD can provide a first-principles determination of the
hadronic contribution to the running of QED coupling. A target
precision  $\lesssim 1.5\%$ would be needed in order to reach a
comparable precision than the dispersive approach at intermediate
scales of a few GeV.

We report about the status of our ongoing calculation of the running
of the electromagnetic
coupling~\cite{Francis:2014yga,Horch:2013lla,Herdoiza:2014jta} and of
its extension to the study of the running of the weak mixing angle.


\section{Hadronic Contributions to the Running of the QED Coupling}
\label{sec:alp}

The running of the QED coupling $ \alpha(Q^2)$ can be written in the
following way,
\begin{equation}\label{eq:aldef}
  \alpha(Q^2)\ =\ \frac{\alpha}{1- \Delta
    \alpha_{\rm QED}(Q^2)} \,,
\end{equation}
where $\Delta \alpha_{\rm QED}(Q^2)$ is determined from the subtracted
vacuum polarisation function (VPF), $\widehat{\Pi}(Q^2) =
\Pi(Q^2)-\Pi(0)$. The hadronic contribution to the running
therefore takes
the following form,
\begin{equation}\label{eq:Ddef}
  \Delta\alpha_{\rm QED}^{\rm had}(\hat Q^2)
  \ =\  4 \pi \alpha \, \widehat{\Pi}(\hat Q^2)\,,
\end{equation}
where the hadronic VPF can be determined from the vector correlation
function involving the electromagnetic current,
\begin{equation}
  J_\mu(x) \ = \ \mathlarger{\mathlarger{‎‎\sum}}_{{\rm f}=u,\,d,\,s,\,c,\dots} \, Q_{\rm f} \, \psibar_{\rm
    f}(x)\gamma_\mu \psi_{\rm f}(x) \,.
  \label{eq:current}
\end{equation}
In eq.~(\ref{eq:current}), $Q_{\rm f}$ refers to the electric charge of
the quark flavour ${\rm f}$. In order to circumvent the requirement of
a direct determination of the subtraction term $\Pi(0)$, it is
possible to determine $\Delta\alpha_{\rm QED}^{\rm had}(\hat Q^2)$
from a lattice QCD computation of the Adler function,
\begin{equation}
  D(\hat Q^2)
  \ = \ 12\pi^2 \,\hat Q^2\,\frac{d\,\Pi\,(\hat Q^2)}{d\hat Q^2}
  \ = \ \frac{3\pi}{\alpha}\,\hat Q^2\,\frac{d}{d\hat Q^2}\Delta
  \alpha_{\rm QED}^{\mathrm{had}}(\hat Q^2)\,,
  \label{eq:defadler}
\end{equation}
where, $\hat{Q}_\mu={2}/{a} \,\sin\left({a\,Q_\mu}/{2}\right)$ is the
lattice momentum.

We update our study of the quark-connected contribution to $D(\hat
Q^2)$ by including eleven CLS ensembles with two flavours of O($a$)
improved Wilson fermions at three values of the lattice spacing and
pion masses down to $190$\,MeV fulfilling the condition, $M_{\rm
  PS}\,L \gtrsim 4$. We consider the valence contributions from $u$,
$d$, $s$ and $c$ quarks and profit from the use of partially twisted
boundary conditions~\cite{DellaMorte:2011aa} to construct the Adler
function from the numerical derivative of the
VPF~\cite{Francis:2014yga,Horch:2013lla,DellaMorte:2014rta,Francis:2014dta}. The
present study also benefits from a significant increase of statistics,
an updated estimate of the bare parameters corresponding to the
physical strange and charm quark masses and a refined implementation
of the method to compute the Adler function.

\begin{figure}[t!]
  \centering \subfigure[\label{fig:DMud_a}]{
    \includegraphics[height=0.41\linewidth]{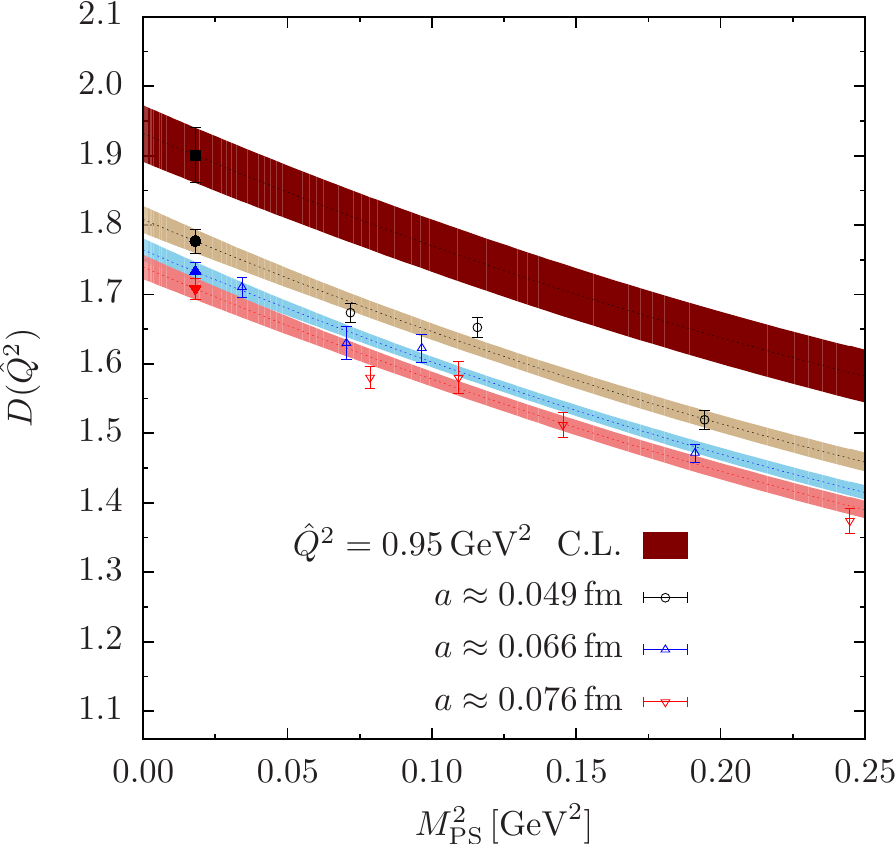}}
  \qquad \subfigure[\label{fig:DMud_b}]{
    \includegraphics[height=0.41\linewidth]{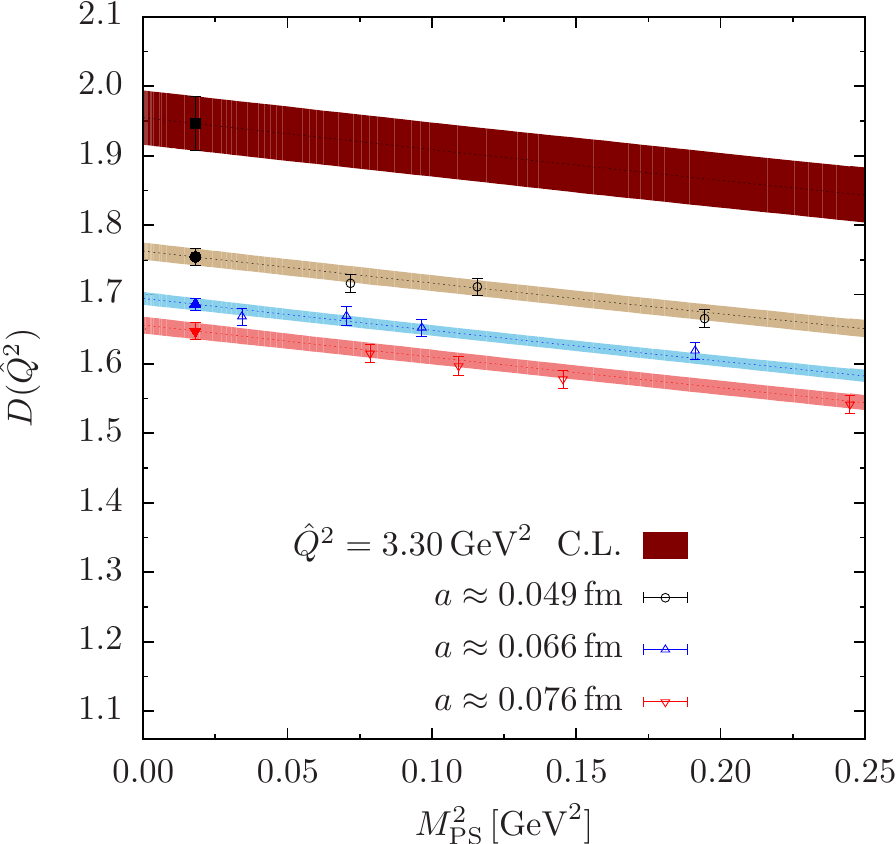}}
  \caption{Dependence on the pion mass squared $M_{\rm PS}^2$ of the
    $(u,d)$ contribution to the Adler function at two values of $\hat
    Q^2$, (a) ${\hat Q}^2=0.95\,{\rm GeV}^2$ and (b) ${\hat
      Q}^2=3.30\,{\rm GeV}^2$. The coloured bands represent a fit
    including a quadratic form in $M_{\rm PS}^2$ to parametrise the
    mass dependence. We observe the suppression of the mass effects
    when increasing the momentum transfer. The dark upper band,
    labelled `C.L', is the continuum limit estimate based on an
    extrapolation with O$(a)$ terms. As expected, we observe that
    lattice artefacts are more pronounced at large $Q^2$. The
    uncorrelated fit combining all the ensembles in the $(u,d)$ sector
    has a $\chi^2/{\rm dof}=0.93$.}
  \label{fig:DMud}
\end{figure}
The lattice determination of the Adler function from the various
ensembles is described by a fit ansatz that parametrises the momentum
dependence, the lattice artefacts and the chiral extrapolation. The
three valence contributions from $(u,d)$, $s$ and $c$ quarks are
fitted independently. In this work, we focus on a momentum interval,
$Q^2 \in [0.5\,;\,4.5]\,{\rm GeV}^2$, that allows one to reduce the
systematic uncertainties associated with both long and short distance
effects. In Fig.~\ref{fig:DMud} we illustrate the result of a fit of
the $(u,d)$ contribution to the Adler function. The pion mass
dependence at two fixed values of $Q^2$ is shown.

\begin{figure}[t!]
  \centering
  \subfigure[\label{fig:DMs}]{
    \includegraphics[height=0.41\linewidth]{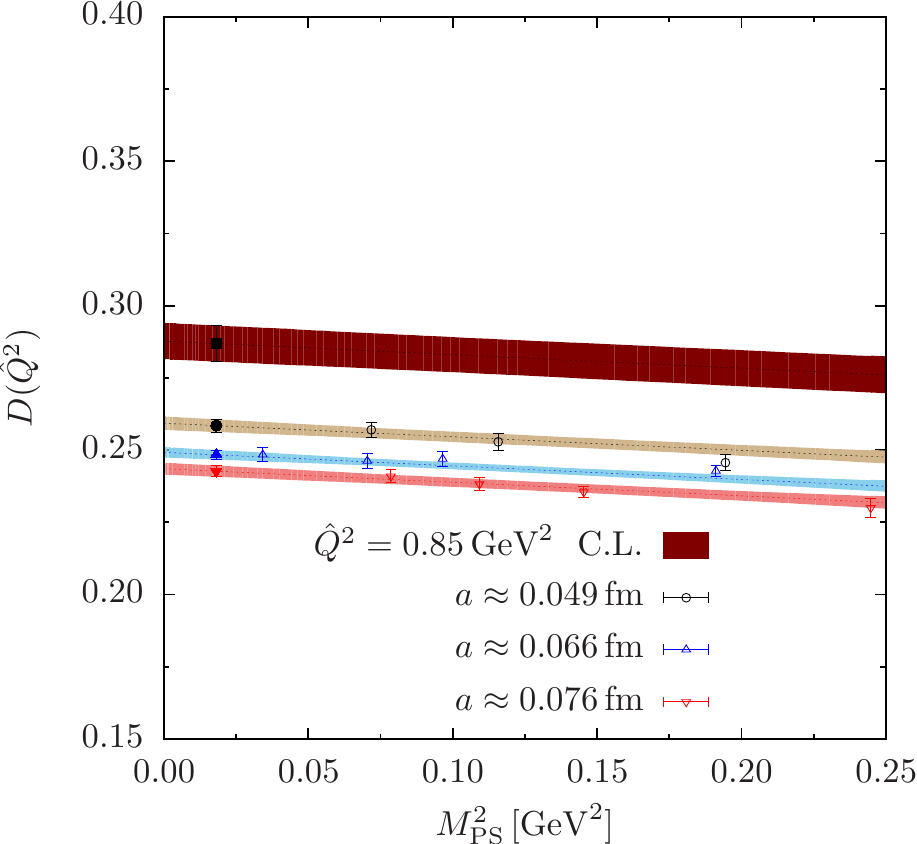}}
  \qquad
  \subfigure[\label{fig:DMc}]{
    \includegraphics[height=0.41\linewidth]{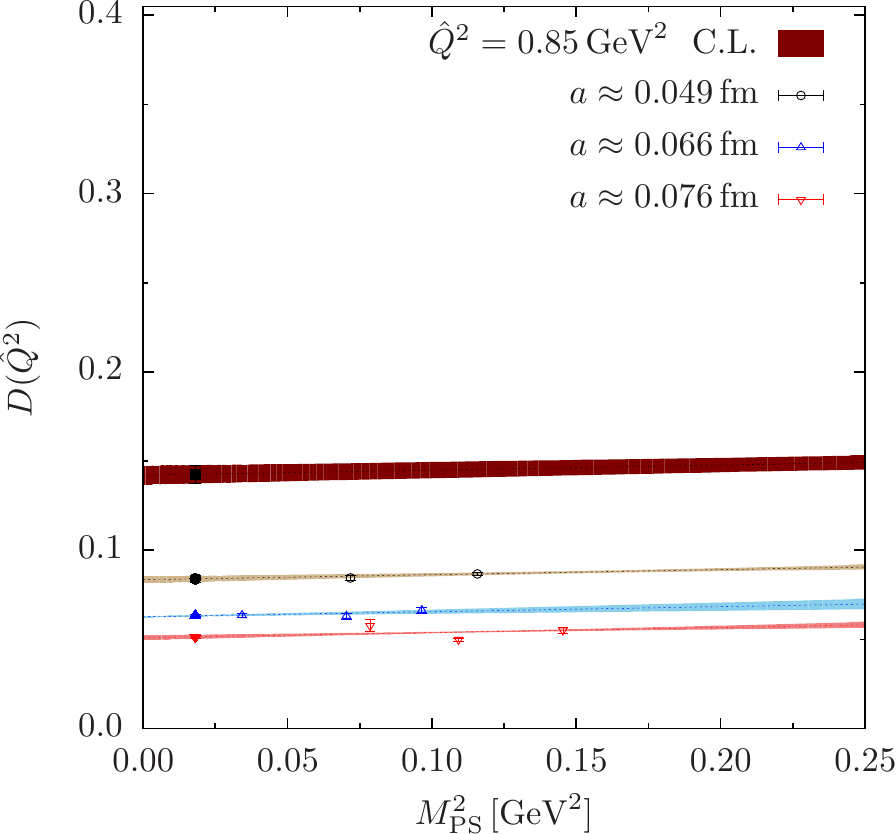}}
  \caption{Pion mass dependence of the (a) strange and (b) charm
    valence contributions to the Adler function at fixed momentum
    transfer, ${\hat Q}^2=0.85\,{\rm GeV}^2$. The $s$ and $c$
    contributions to the Adler function are suppressed with respect to
    the $(u,d)$ case shown in Fig.~\protect\ref{fig:DMud}. Furthermore, since
    the light-quark effects arise only from sea quarks, a mild pion
    mass dependence is observed in both (a) and (b) panels. The
    relative effect from lattice artefacts increases for heavier
    valence quark masses.}
  \label{fig:DMsc}
\end{figure}

In the strange and
charm valence sectors, the relative contribution to the Adler function
is suppressed with respect to the one from the light quark
sector. However, it becomes more and more relevant at larger energy
scales and has, in any case, a significant effect given the present
level of precision.  Fig.~\ref{fig:DMsc} shows the pion mass
dependence of the Adler function in the $s$ and $c$ valence sectors.

\begin{figure}[t!]
  \centering
  \subfigure[\label{fig:DalphaQa}]{
    \includegraphics[height=0.41\linewidth]{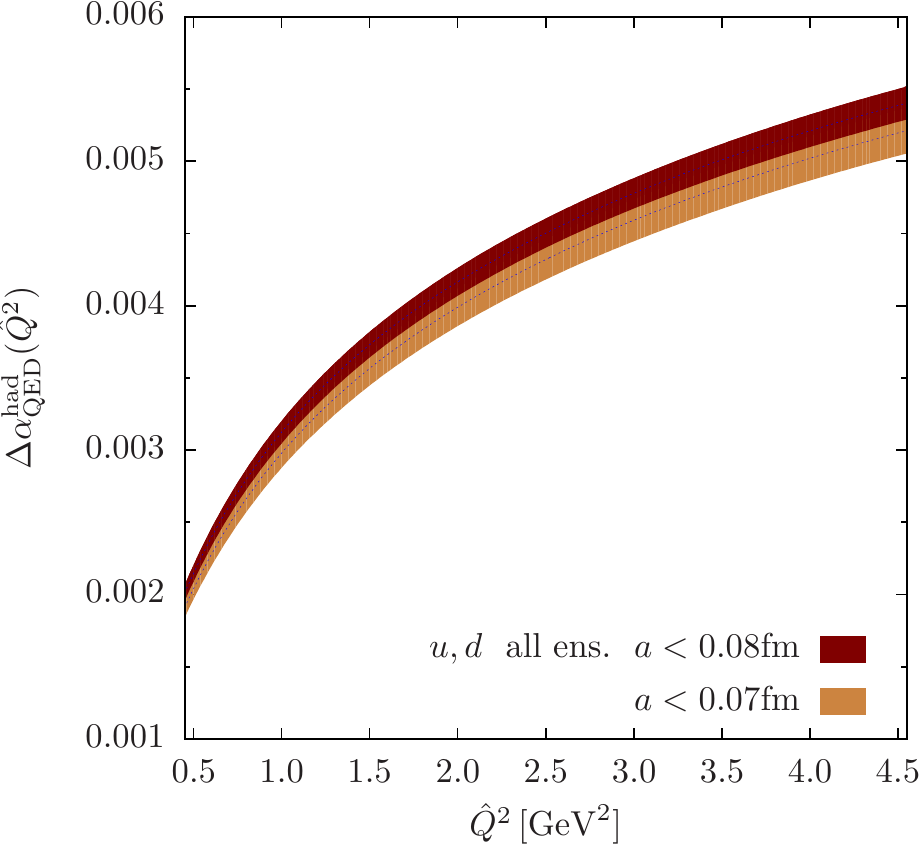}}
  \qquad
  \subfigure[\label{fig:DalphaQb}]{
    \includegraphics[height=0.41\linewidth]{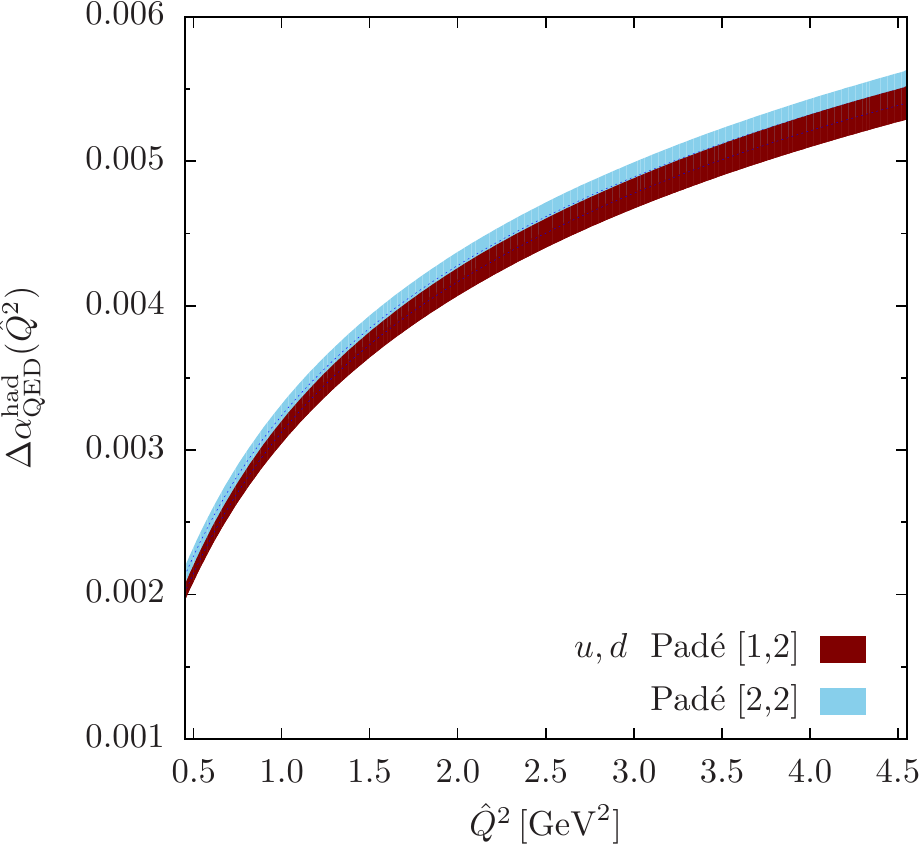}}
  \caption{Momentum dependence of the $(u,d)$ quark-connected
    contribution to the running of the QED coupling $\daqhq$ in the
    continuum and at the physical point. (a) The dark band shows the
    result from a continuum limit extrapolation with O($a$) terms
    including the complete set of ensembles while for the lighter
    band, the ensembles from the coarser lattice spacing were
    excluded. (b) Comparison of the use of two different orders of the
    Pad\'e approximants to parametrise the momentum dependence of
    $D(\hat Q^2)$.}
  \label{fig:DalphaQ}
\end{figure}

The continuum result for the $D(Q^2)$ at the physical point can be
used to derive $\daqhq$ through eq.~(\ref{eq:defadler}). We explore the
systematic effects in our determination of $\daqhq$ by considering
various fit forms to describe the lattice artefacts, the momentum and
the mass dependence of the Adler function. We also monitor the effect
of removing the coarsest lattice spacing or the heavier ensembles to
assess the systematic uncertainties. In Fig.~\ref{fig:DalphaQ} we
illustrate two of the major contributions to systematic effects in
$\daqhq$ that have so far been analysed.

\begin{figure}[t!]
  \centering
  \subfigure[\label{fig:DQsuma}]{
    \includegraphics[height=0.41\linewidth]{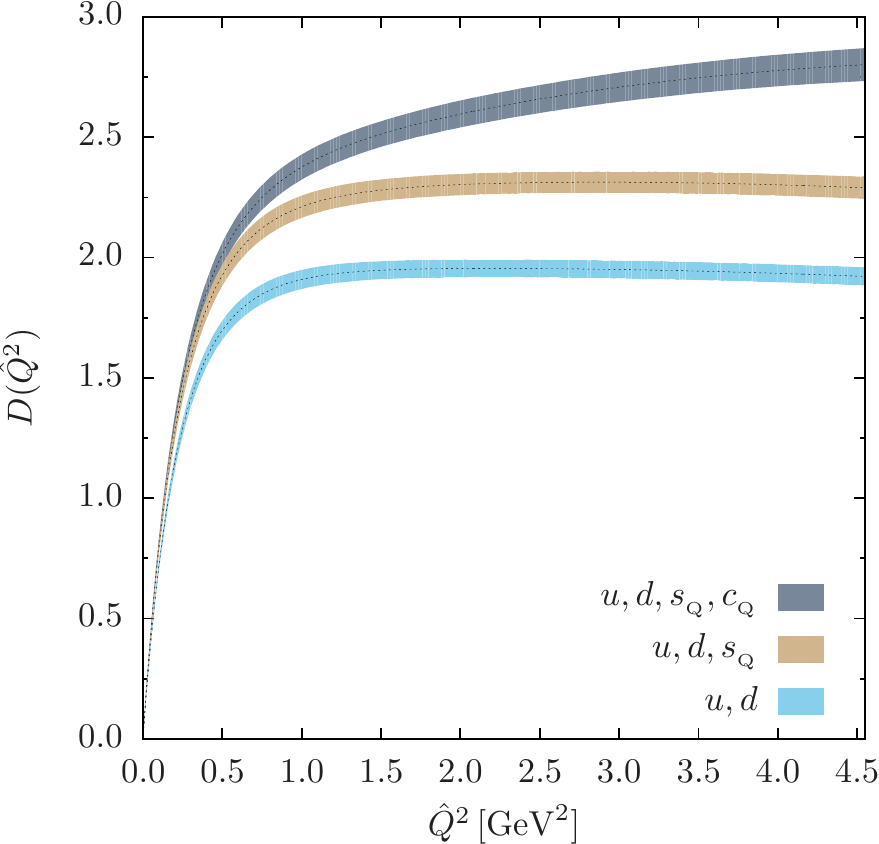}}
  \qquad
  \subfigure[\label{fig:DQsumb}]{
    \includegraphics[height=0.41\linewidth]{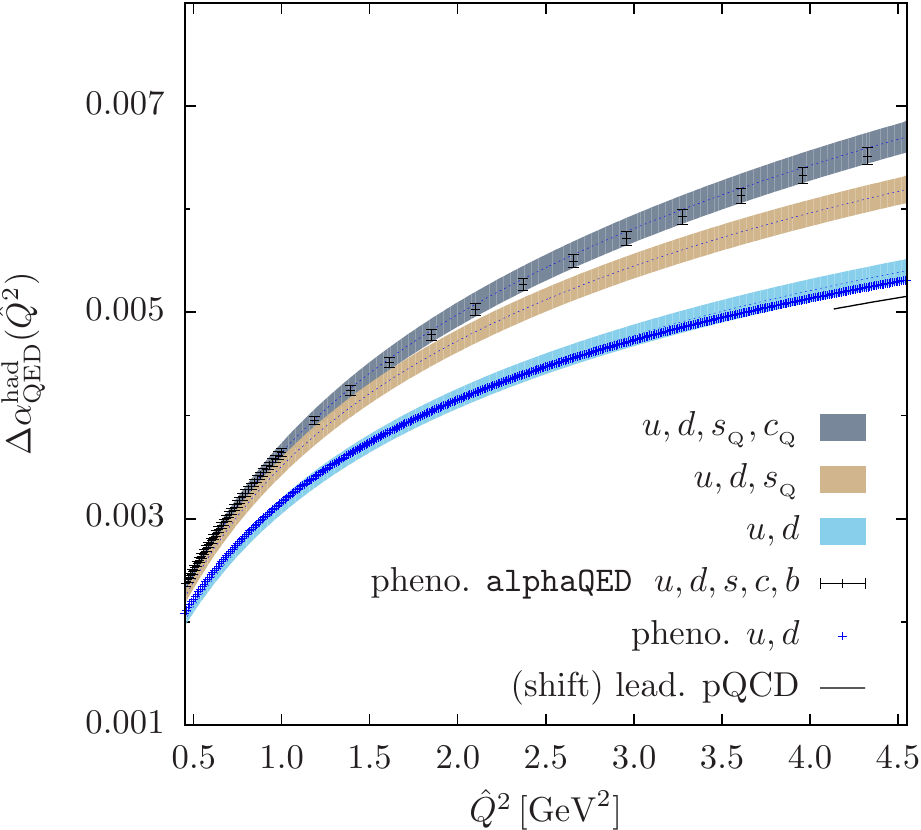}}
  \caption{(a) Contributions to $D(\hat Q^2)$ from $(u,d)$ and from
    partially quenched strange $s_Q$ and charm $c_Q$ quark flavours in
    the continuum and at the physical point. (b) Hadronic contribution
    to the $\daqhq$ from $(u,d)$, $s_Q$ and $c_Q$ quarks. The
    five-flavour result from a dispersive approach implemented through
    the \texttt{alphaQED} package~\cite{Jegerlehner:2011mw,pqcdadler}
    is denoted by the black symbols. The $(u,d)$ contribution is
    compared to the phenomenological model of
    ref.~\cite{Bernecker:2011gh}. The LO perturbative QCD result for
    $(u,d)$ quark flavours is represented by a short black curve that
    was shifted vertically to improve the visibility.  }
  \label{fig:DQsum}
\end{figure}

The effect of summing up the various flavour contributions to the
Adler function and the running of the QED coupling is shown in
Fig.~\ref{fig:DQsum}. We observe that in the $Q^2$ interval, $Q^2 \in
[0.5,\,4.5]\,{\rm GeV}^2$, the lattice QCD determination of $\daqhq$
from $u$, $d$, $s_Q$ and $c_Q$ quarks is compatible with the five
flavour result from a dispersive
approach~\cite{Jegerlehner:2011mw}. At present, however, the lattice
results have larger uncertainties than the dispersive method. Our
preliminary results also agree with a recent study using Wilson
twisted mass fermions~\cite{Burger:2015lqa}.

\section{Hadronic Contributions to the Running of $\sin^2\theta_W$}
\label{sec:sin2tw}

The weak mixing angle $\sin^2\theta_W$ controls the $\gamma-Z$ mixing
and  provides a relation among the coupling constants of the
electroweak theory. The value of $\sin^2\theta_W$ at the $Z$-pole has
been determined with good precision by the LEP experiments and is
heavily constrained by the global fit of the SM in the electroweak
sector~\cite{Baak:2014ora}.

A number of ongoing and future low-energy
experiments~\cite{Erler:2013xha} aim at determining $\sin^2\theta_W$
at energy scales below $3\,{\rm GeV}$ where non-perturbative QCD
effects start to be relevant. These hadronic effects cannot be
accommodated by a straightforward application of a dispersive approach
due to the difficulty in isolating the contributions from up- and
down-type quarks. On the other hand, such a flavour separation is
naturally provided by a lattice QCD calculation. The hadronic
contribution to the running of the weak mixing angle, $\Delta^{\rm
  had}\sin^2\theta_W({Q}^2)$, can then be related to $\daqhq$ once an
independent input for the value of the SU(2)$_L$ coupling $\alpha_2$
at $Q^2=0$ is employed~\cite{Kumar:2013yoa}.

\begin{figure}[t!]
  \centering
  \includegraphics[height=0.5\linewidth]{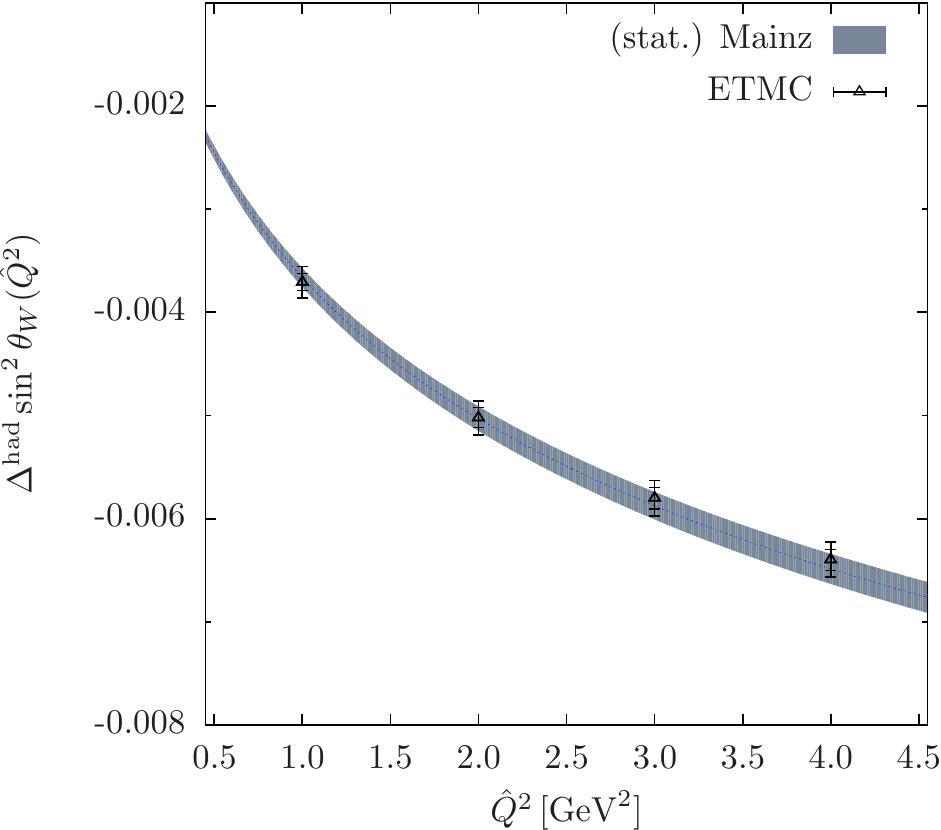}
  \caption{Quark-connected contribution to the running of the weak
    mixing angle. The coloured band represents our preliminary results
    for the $u$, $d$, $s$ and $c$ valence quark contribution to
    $\Delta^{\rm had}\sin^2\theta_W({Q}^2)$. The empty triangles refer
    to the recent lattice computation with Wilson twisted mass
    fermions~\cite{Burger:2015lqa}.}
  \label{fig:sin2twQ}
\end{figure}

A lattice computation of the quark-connected contribution to the
running of $\sin^2\theta_W$ was presented
in~Ref.~\cite{Burger:2015lqa} while the effect of including the
disconnected contributions was discussed in
Ref.~\cite{guelpers:lat2015}. By extending the study of the running of
the QED coupling presented in the previous section, we obtain a
determination of the connected contribution from $u$, $d$, $s$ and $c$
valence quarks to $\Delta^{\rm
  had}\sin^2\theta_W({Q}^2)$. Fig.~\ref{fig:sin2twQ} shows our
preliminary results together with a comparison to the lattice
computation of~Ref.~\cite{Burger:2015lqa}.


\section*{Conclusions}

We have presented the status of our ongoing study of the hadronic
contributions to the running of the electroweak couplings from a
lattice QCD calculation of the Adler function. Our preliminary results
indicate that for the case of the running of the QED coupling, a
further reduction of the systematic effects -- in particular those
related to lattice artefacts -- would be needed to match the precision
of the dispersive approach at intermediate values of the momentum
transfer. The computation of the hadronic contributions to the running
of the weak mixing angle is required to confront the SM prediction to
a number of ongoing low-energy experiments. We refer to
Refs.~\cite{guelpers:lat2015,horch:lat2015} for an account of related
studies presented at this conference concerning the quark-disconnected
contribution to the running of the weak mixing angle and the
leading-order hadronic contribution to the muon $g-2$.


\paragraph*{Acknowledgements}

The calculations were performed on the ``Wilson'' and ``Clover'' HPC
Clusters at the Institute of Nuclear Physics of the University of Mainz.
We thank Dalibor Djukanovic and Christian Seiwerth for technical support.
This work was granted access to the HPC resources of the Gauss Center
for Supercomputing at Forschungzentrum J\"ulich, Germany, made available
within the Distributed European Computing Initiative by the PRACE-2IP,
receiving funding from the European Community's Seventh Framework
Programme (FP7/2007-2013) under grant agreement RI-283493 (project
PRA039).
We are grateful for computer time allocated to project HMZ21 on the
BG/Q JUQUEEN computer at NIC, J\"ulich.
This research has been supported by the DFG in the SFB~1044.
We thank our colleagues from the CLS initiative for sharing
the ensembles used in this work.
G.H. acknowledges support by the Spanish MINECO through the Ram\'on y
Cajal Programme and through the project FPA2012-31686 and by the
Centro de Excelencia Severo Ochoa Program SEV-2012-0249.


\end{document}